\documentclass[12pt]{iopart}
\usepackage{graphicx}
\usepackage{bm}
\def\beq {\begin{equation}}
\def\enq {\end{equation}}
\newcommand{\DIR}{./figures}

\newcommand{\FORCE}{
\begin{figure}[htbp]
\centering
\includegraphics[angle=0,width=8.0cm]{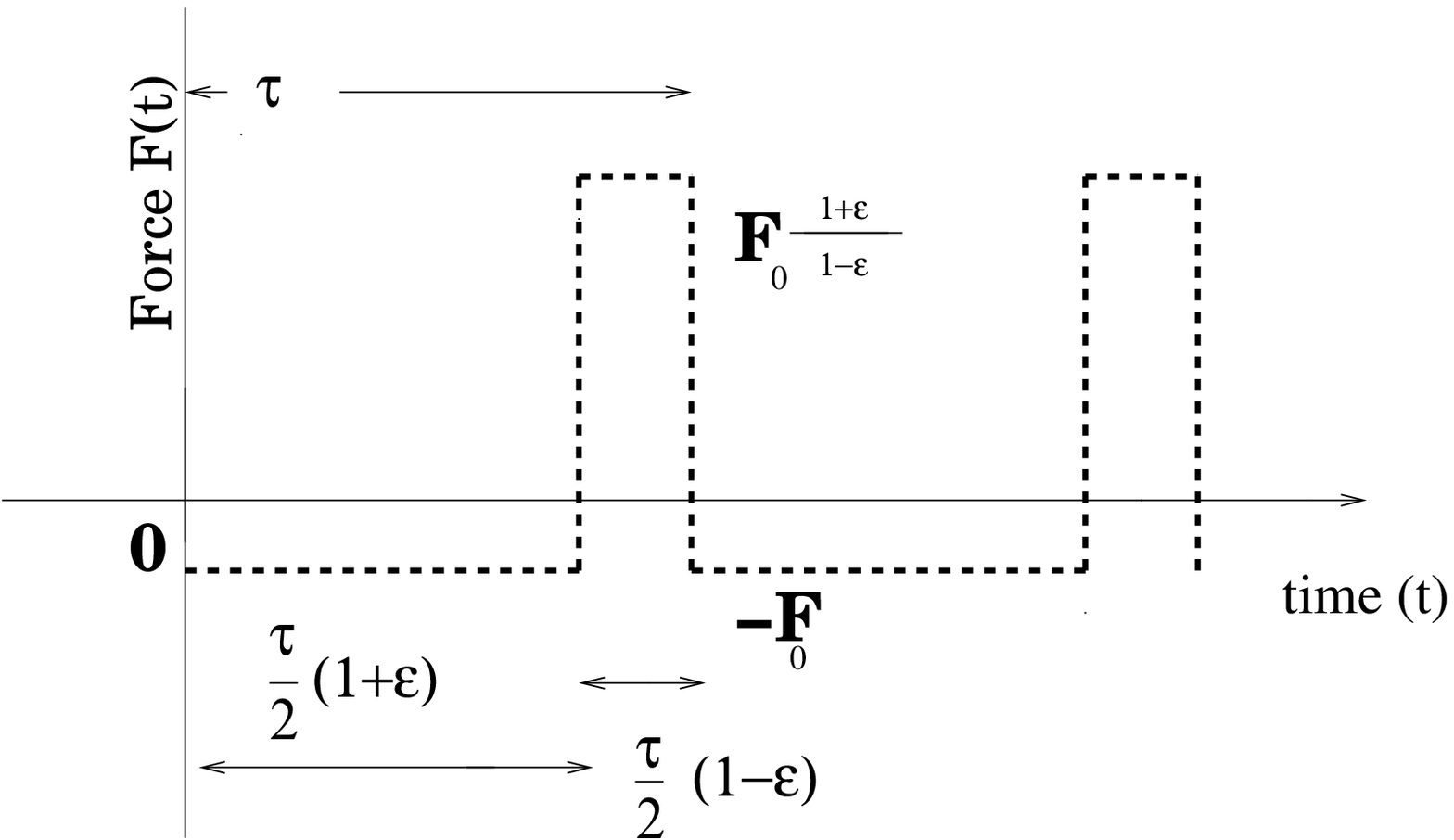}
\caption{Illustration of time asymmetric force for time period $\tau$
and temporal asymmetry factor $\epsilon$}
\label{forceee}
\end{figure}}

\newcommand{\HBFSPL}{
\begin{figure}[htbp]
\begin{center}
\includegraphics[height=3.25in,width=2.0in,angle=270]{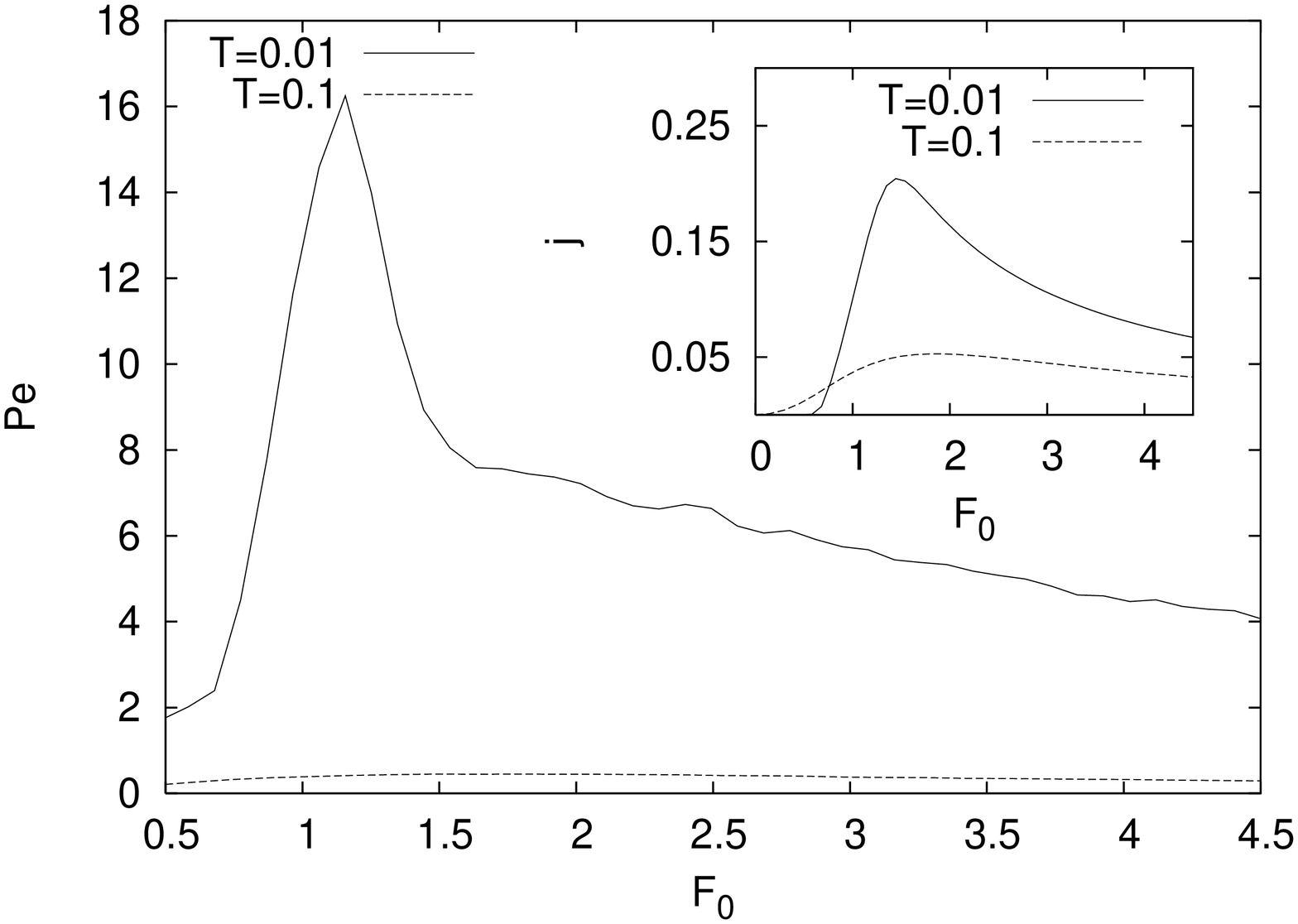}
\caption{Plot of $Pe$ and $j$(inset) versus $F_0$ at
 $T~=~0.01$ and $T=0.1$ in the temporally asymmetric driving case 
for $\omega~=~0.25$(adiabatic limit).}
\label{hbfspl}
\end{center}
\end{figure}}

\newcommand{\HBTSPL}{
\begin{figure}[htbp]
\begin{center}
\includegraphics[height=3.25in,width=2.0in,angle=270]{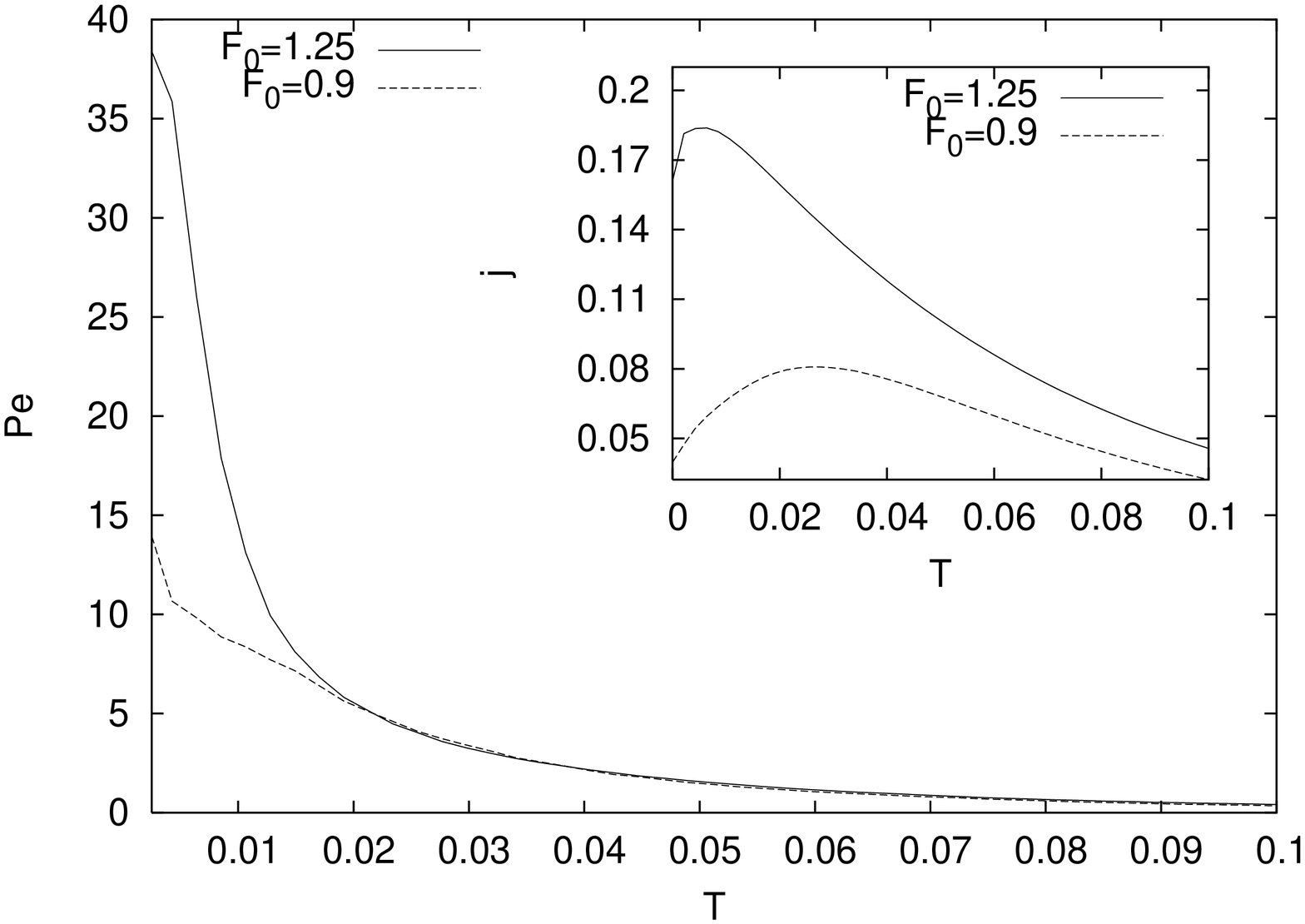}
\caption{Plot of $Pe$ and $j$(inset) versus $T$ at 
$F_0~=~0.9$ and $F_0~=~1.25$ in the temporally asymmetric driving 
case for $\omega~=~0.25$(adiabatic limit).}
\label{hbtspl}
\end{center}
\end{figure}}

\newcommand{\HBF}{
\begin{figure}[htbp]
\begin{center}
\includegraphics[height=3.5in,width=1.9in,angle=270]{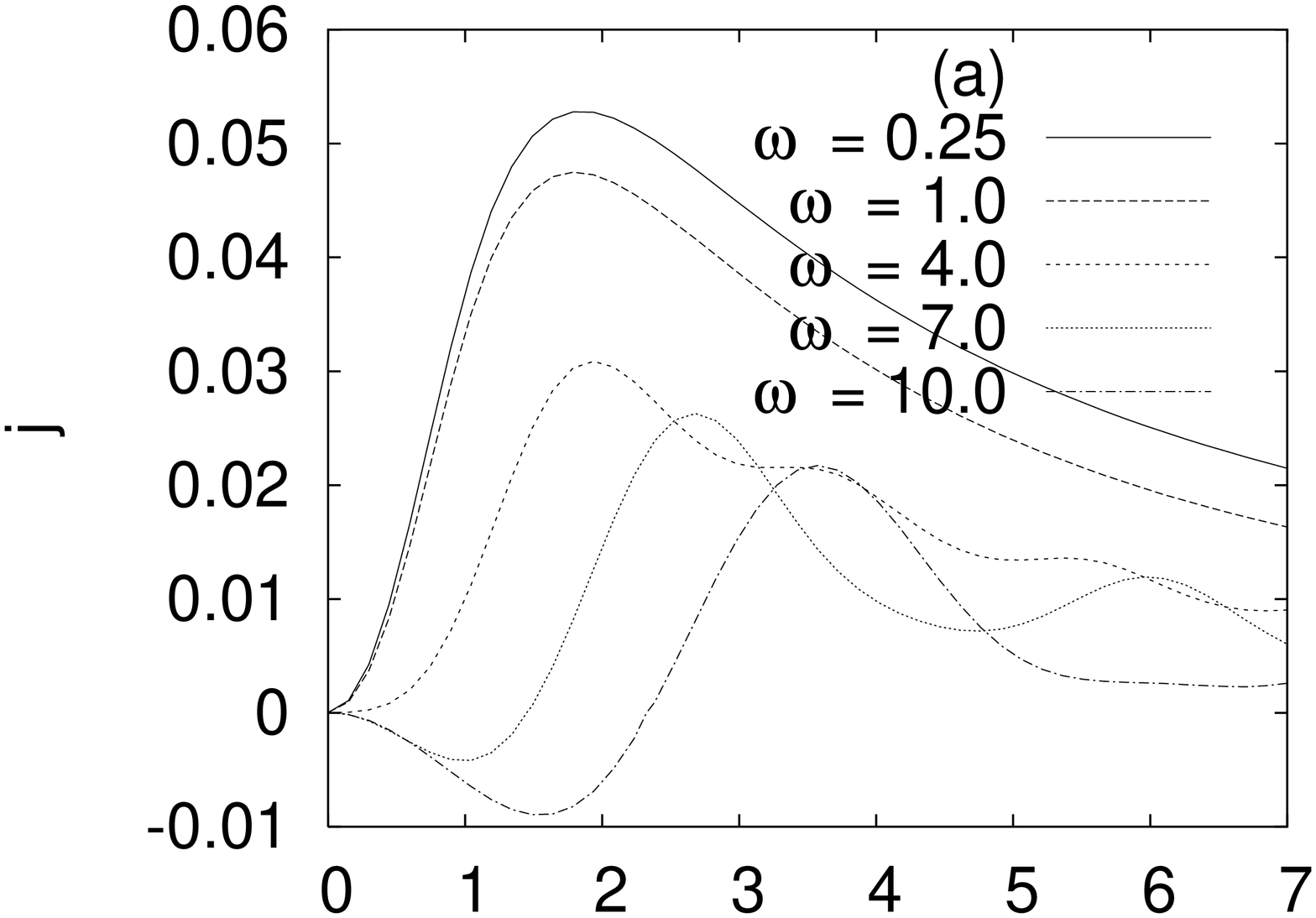}
\includegraphics[height=3.5in,width=1.9in,angle=270]{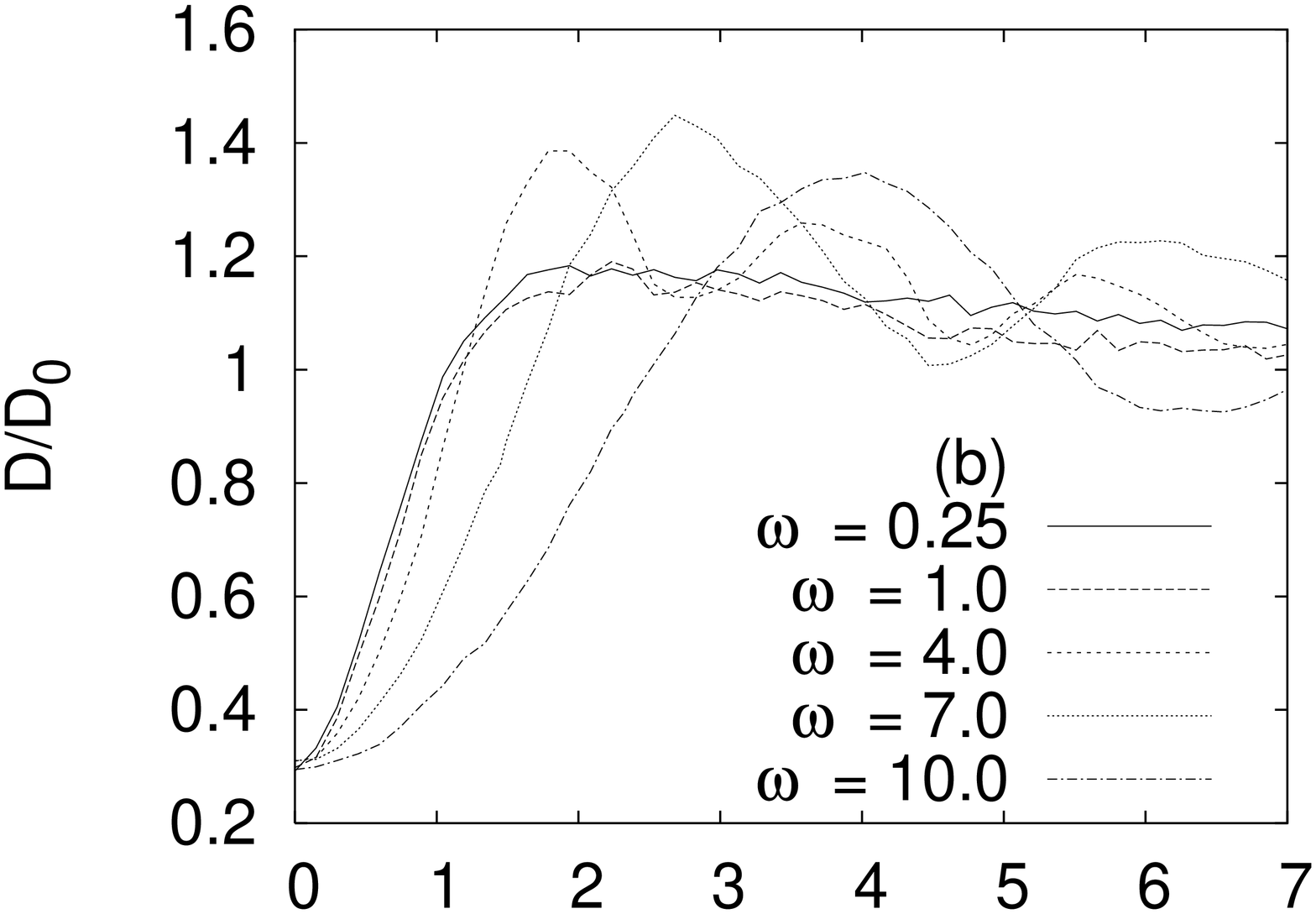}
\includegraphics[height=3.5in,width=1.9in,angle=270]{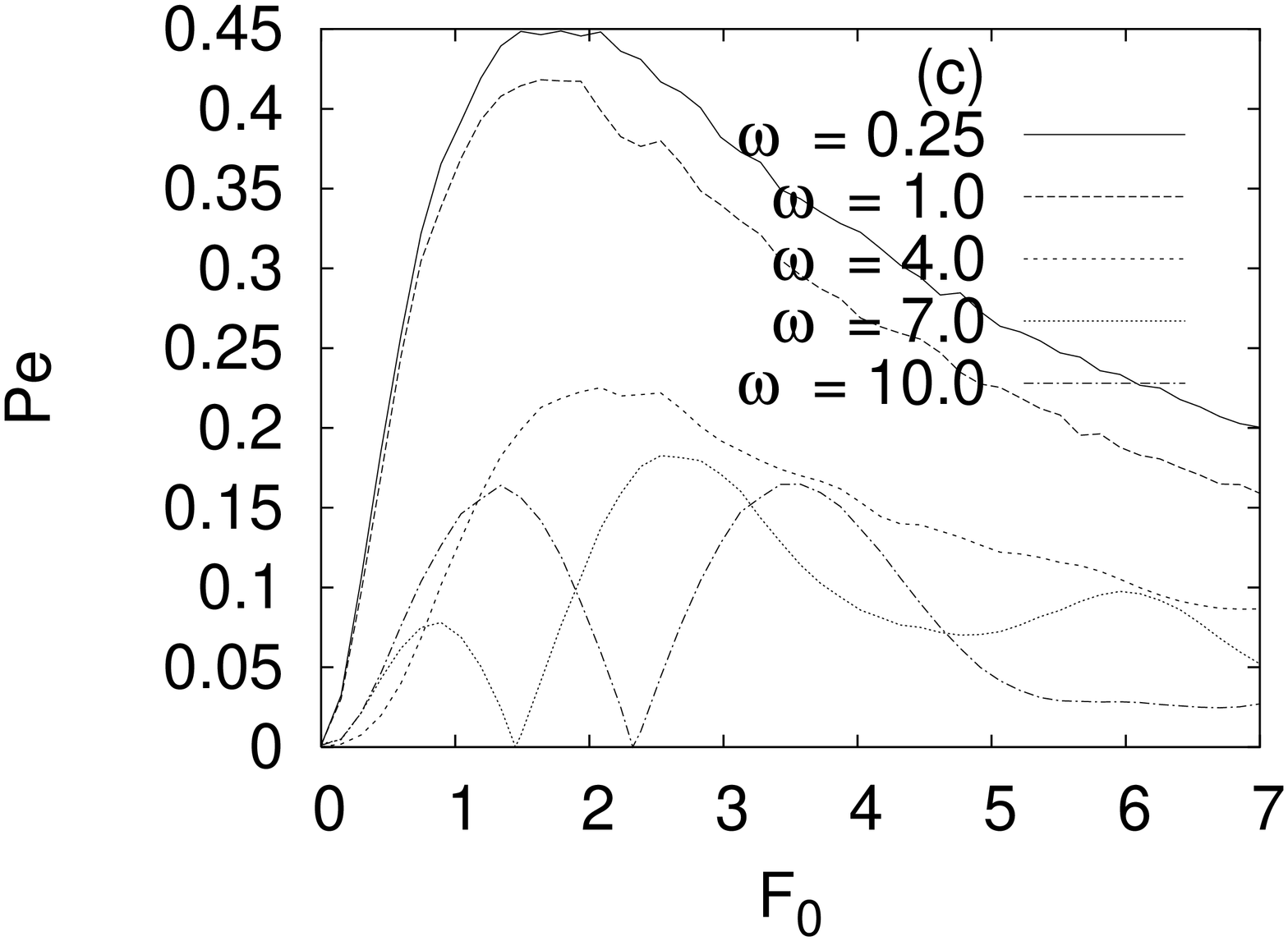}
\caption{Plot of (a) $j$ (b) $D$  and 
(c) $Pe$ (from above) versus 
$F_0$ at $T = 0.1$ for the temporally symmetric driving case.}
\label{hbf}
\end{center}
\end{figure}}

\newcommand{\HBT}{
\begin{figure}[htbp]
\begin{center}
\includegraphics[height=3.5in,width=2.0in,angle=270]{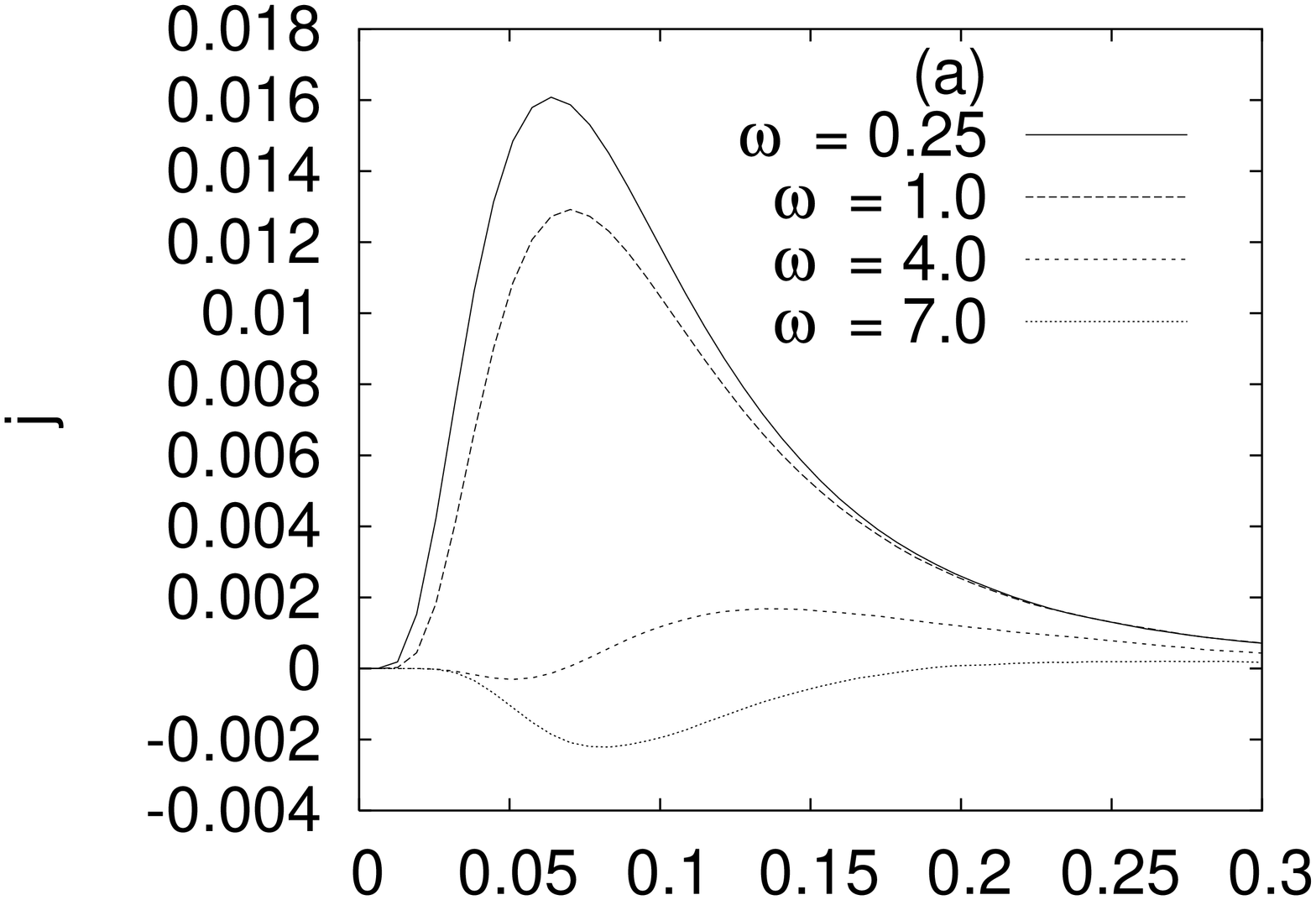}
\includegraphics[height=3.5in,width=2.0in,angle=270]{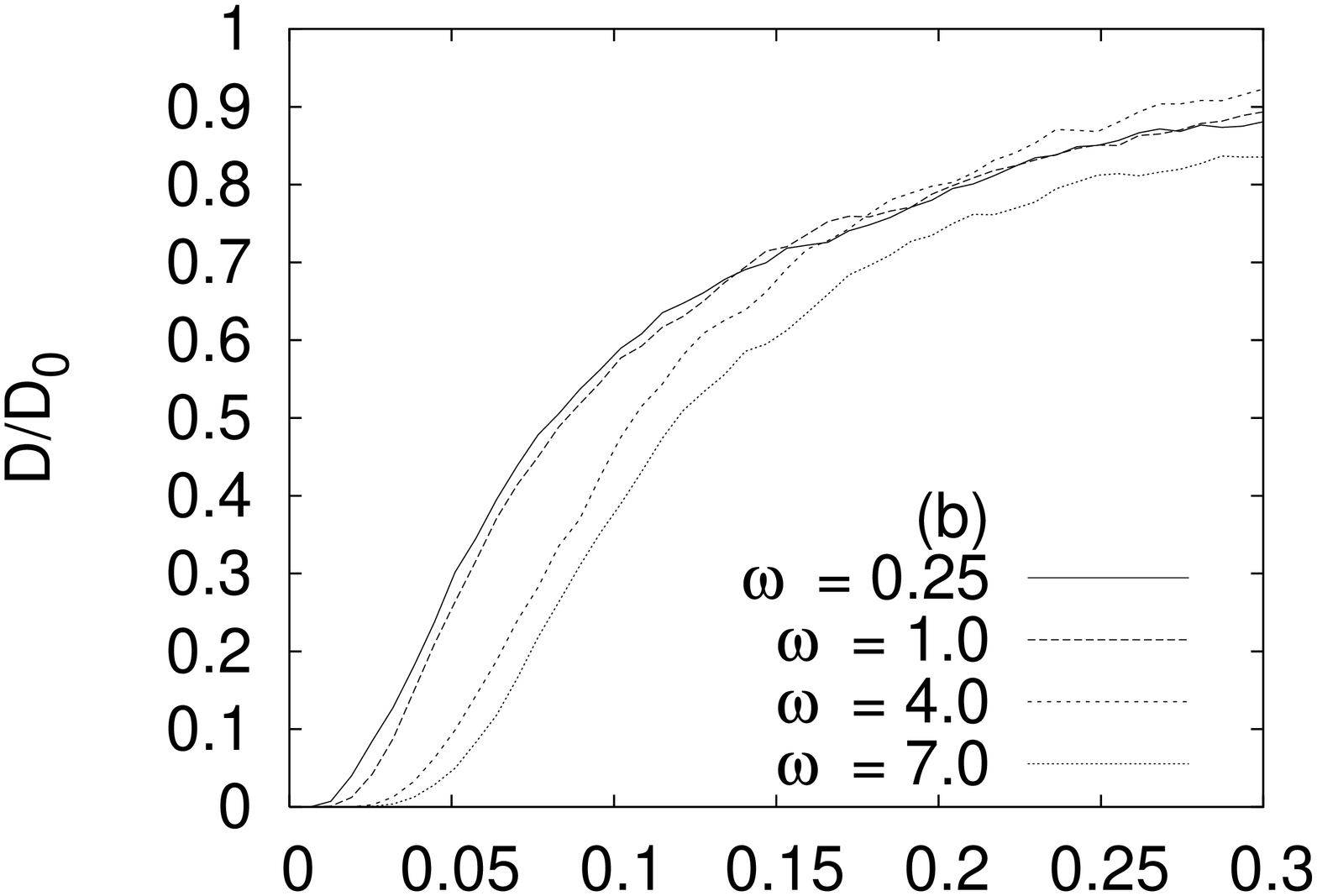}
\includegraphics[height=3.5in,width=2.0in,angle=270]{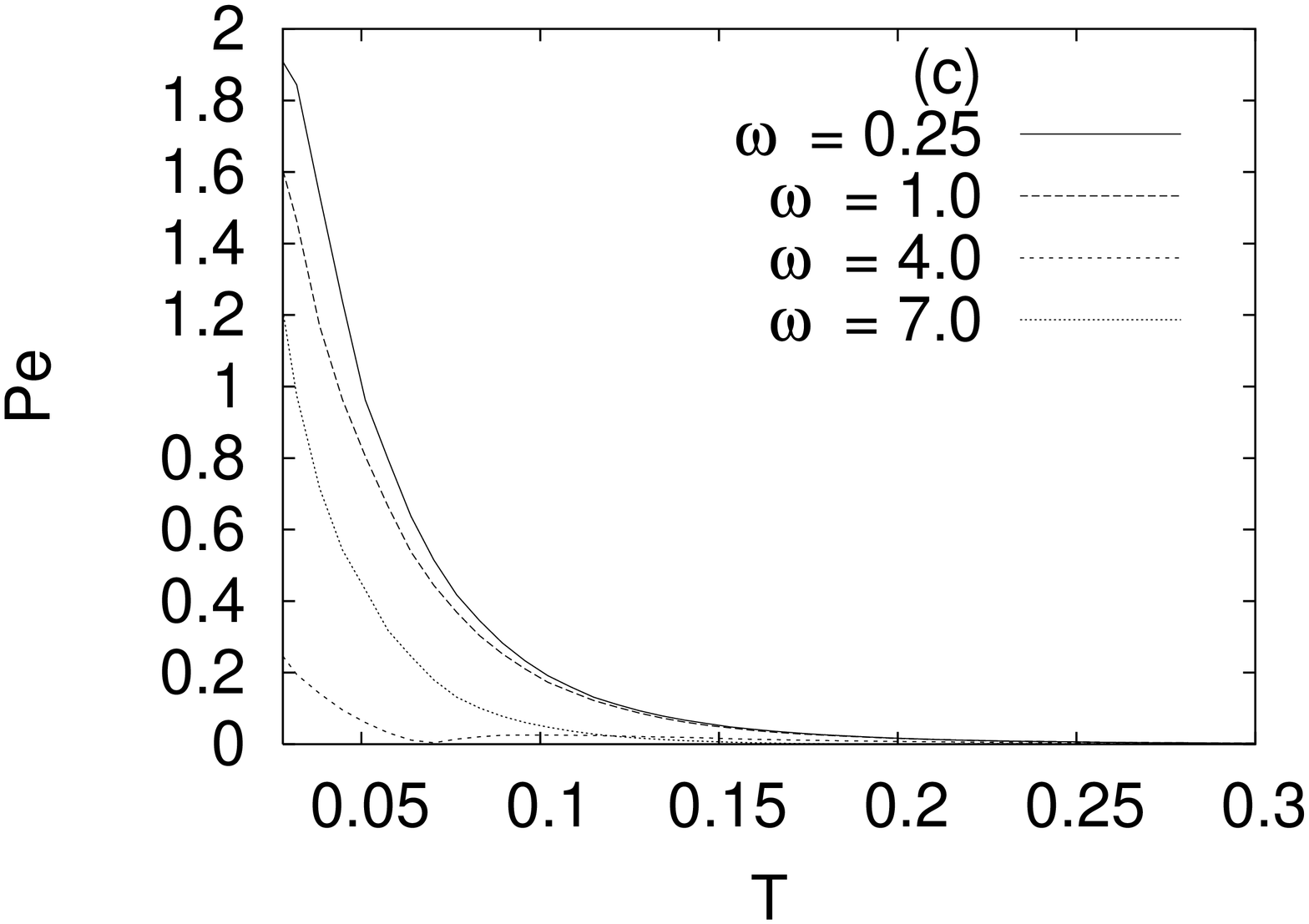}
\caption{Plot of (a) $j$ (b) $D$  and (c) $Pe$ (from above) 
versus $T$ at $F_0 = 0.5$ for the temporally 
symmetric driving case.}
\label{hbt}
\end{center}
\end{figure}}

\newcommand{\SF}{
\begin{figure}[htbp]
\begin{center}
\includegraphics[height=3.5in,width=2.0in,angle=270]{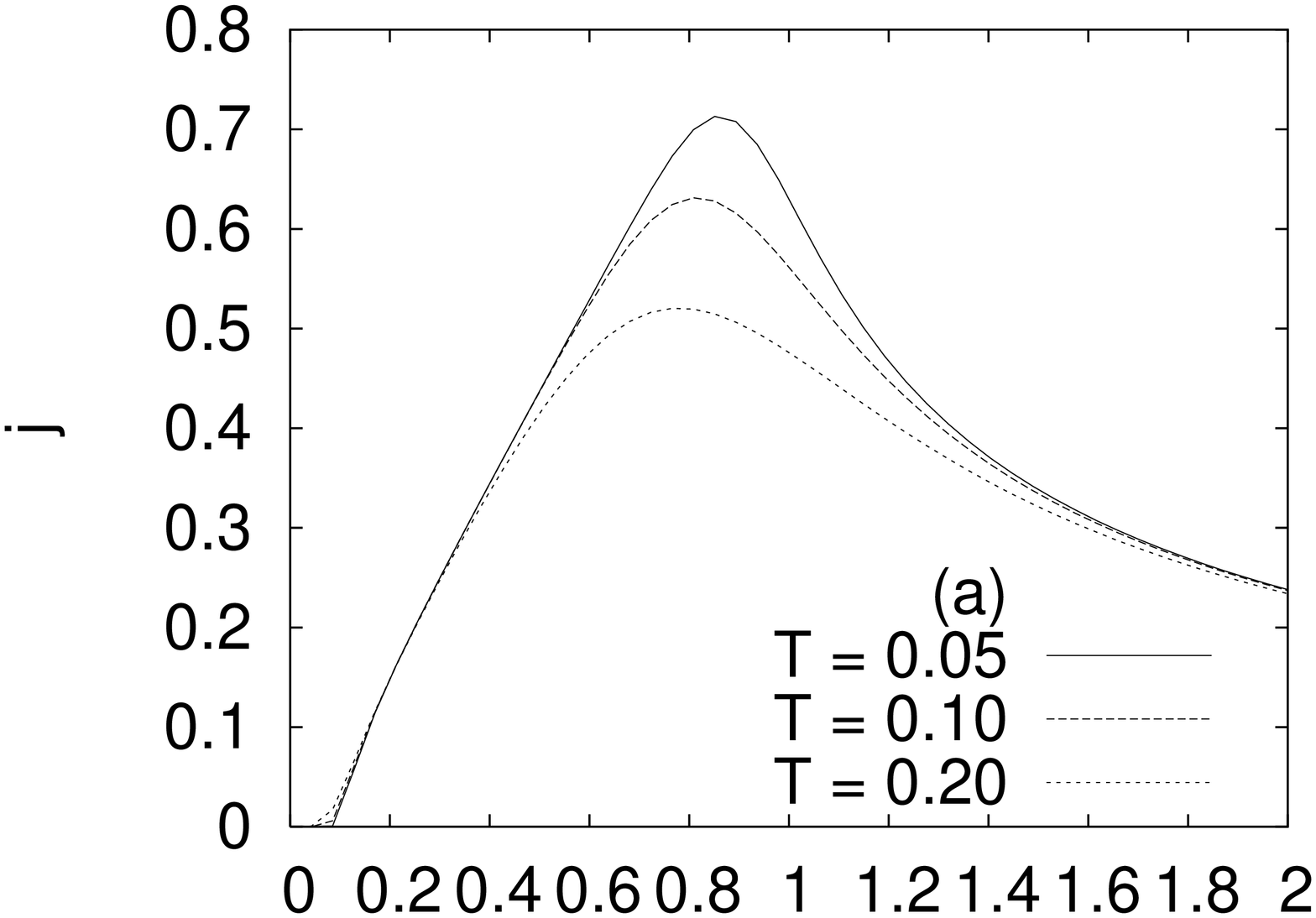}
\includegraphics[height=3.5in,width=2.0in,angle=270]{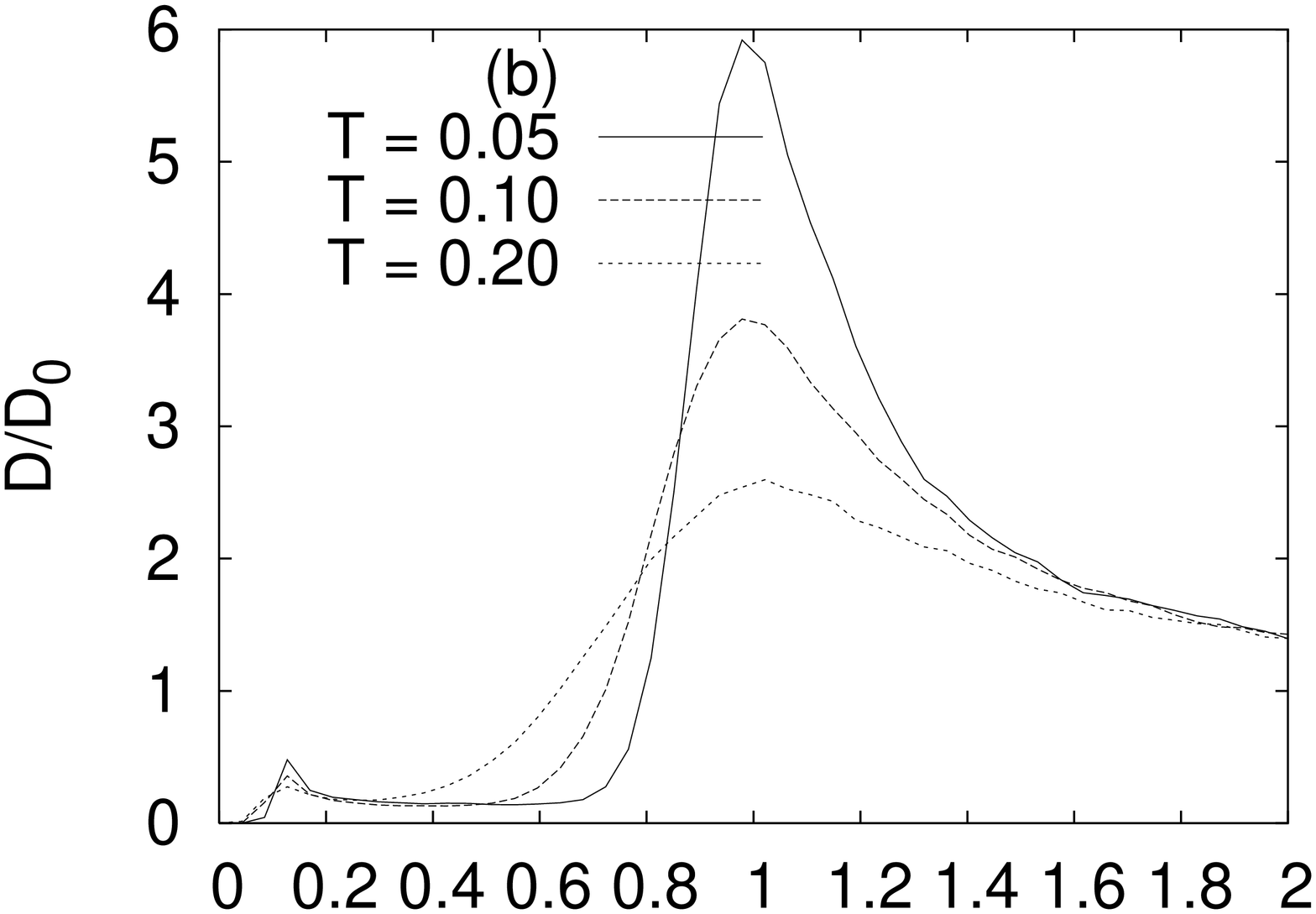}
\includegraphics[height=3.5in,width=2.0in,angle=270]{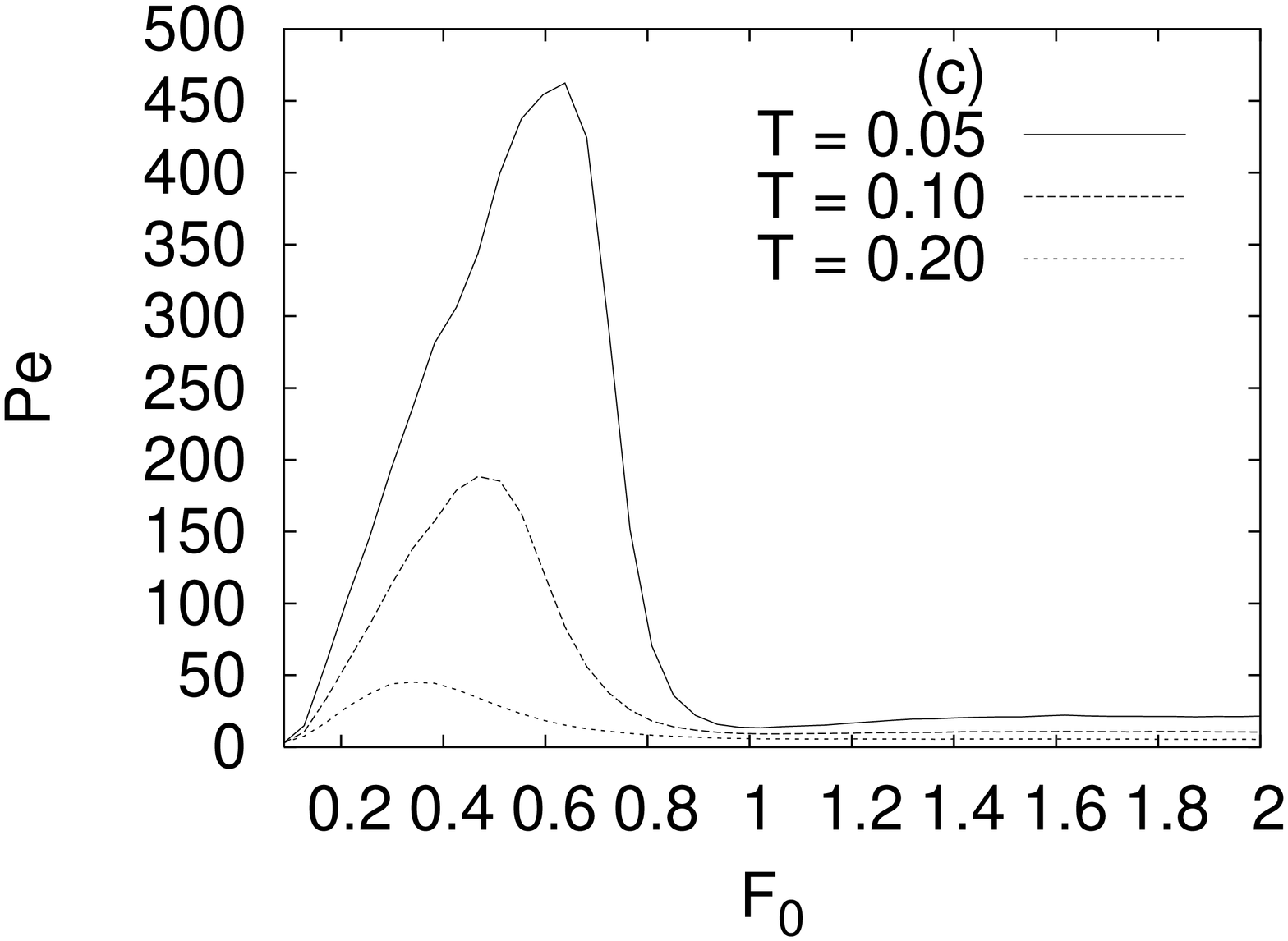}
\caption{Plot of (a) $j$ (b) $D$  and (c) $Pe$ (from above) 
versus $F_0$ at $\epsilon = 0.8$ for various values 
of $T$ in the temporally asymmetric driving case  in the adiabatic 
limit ($\tau=1000$).}
\label{sf}
\end{center}
\end{figure}}

\newcommand{\NONAD}{
\begin{figure}[htbp]
\begin{center}%
\includegraphics[height=3.5in,width=2.0in,angle=270]{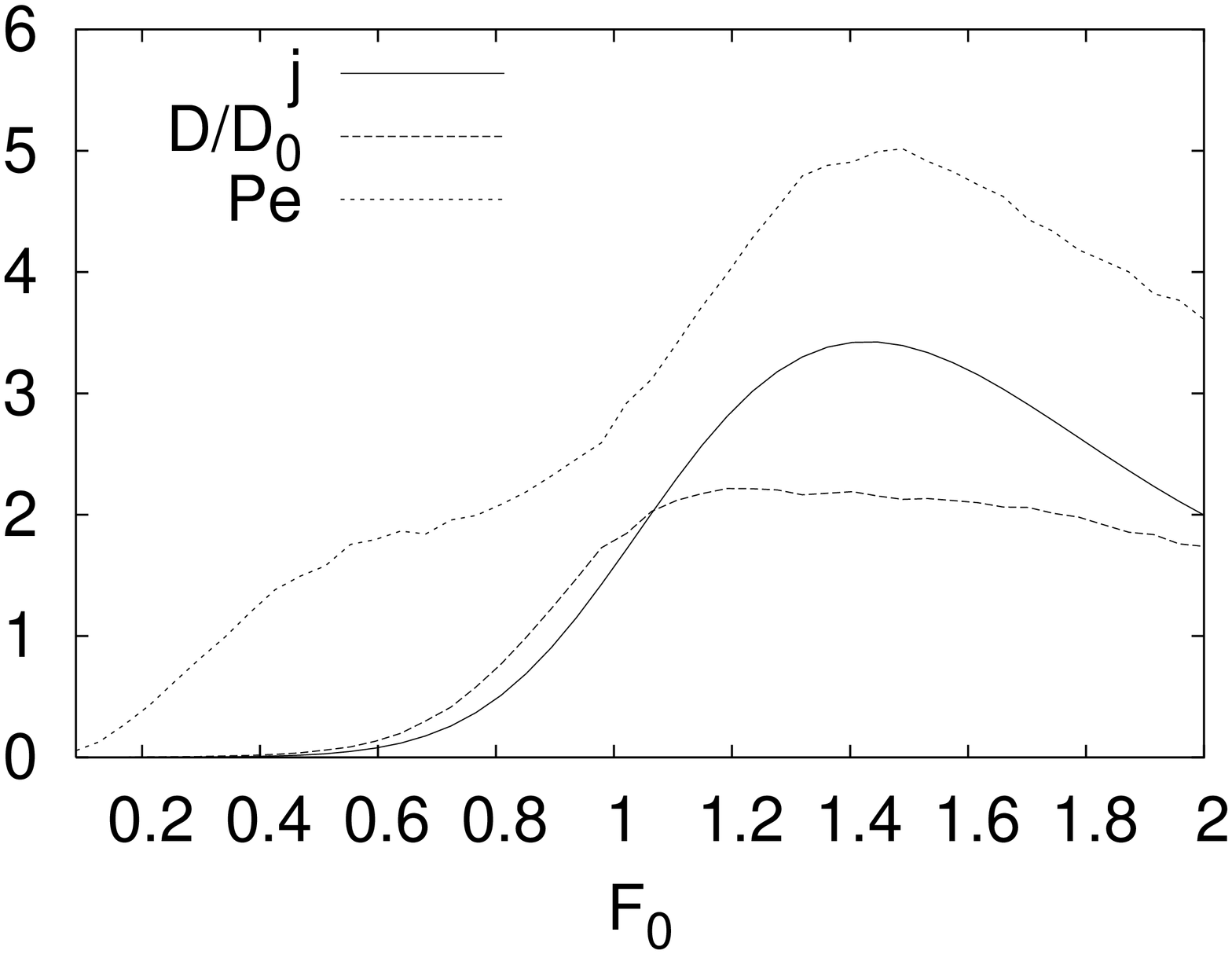}
\caption{Plot of $j$, $D$ and $Pe$ 
 versus $F_0$ at $\epsilon = 0.8$ and $\tau = 5$ (non-adiabatic 
limit) at $T = 0.2$ in the temporally asymmetric driving case. The $j$ 
has been scaled by a factor of 10.}
\label{nonad}
\end{center}
\end{figure}}

\newcommand{\ST}{
\begin{figure}[htbp]
\begin{center}
\includegraphics[height=3.5in,width=2.0in,angle=270]{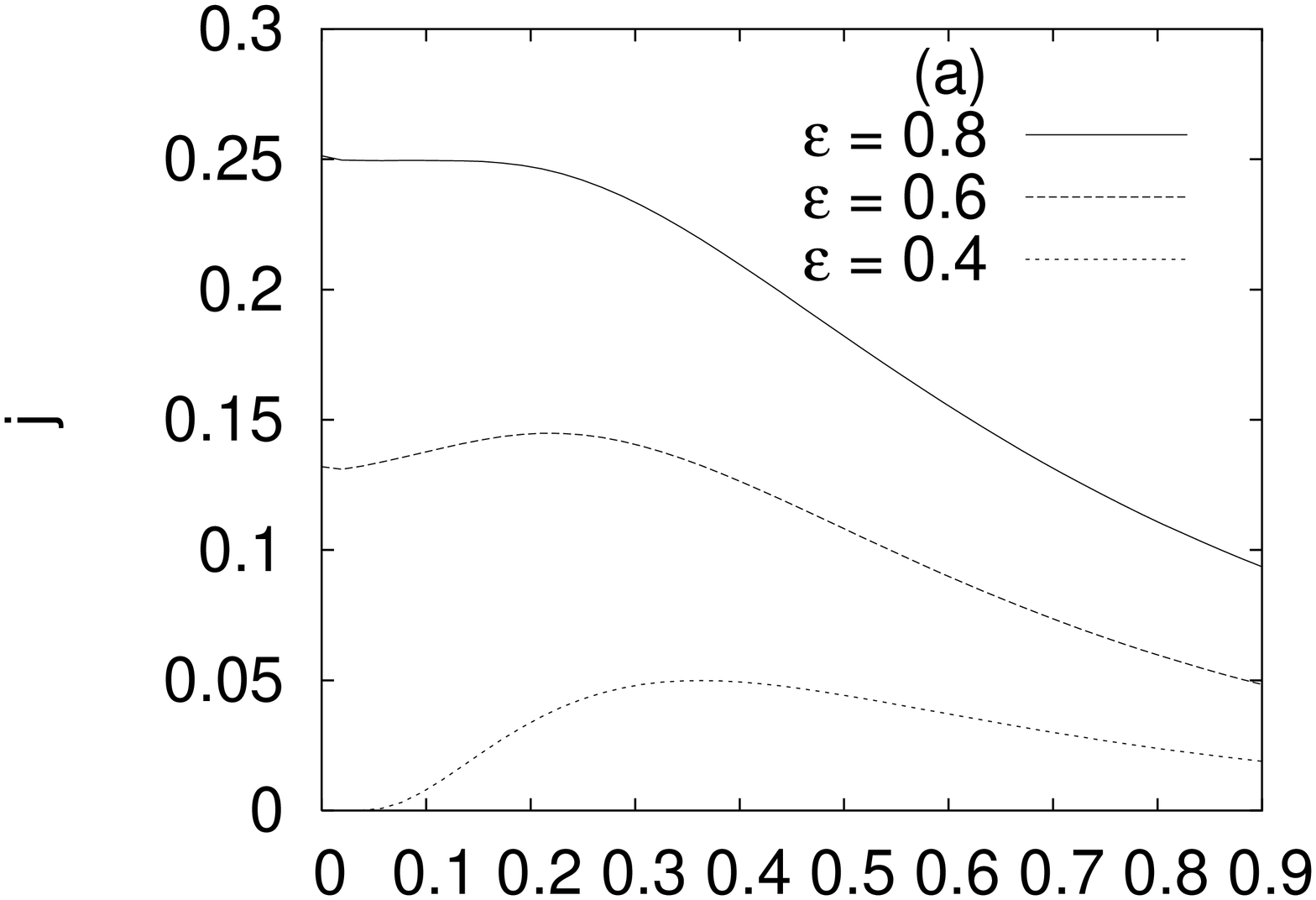}
\includegraphics[height=3.5in,width=2.0in,angle=270]{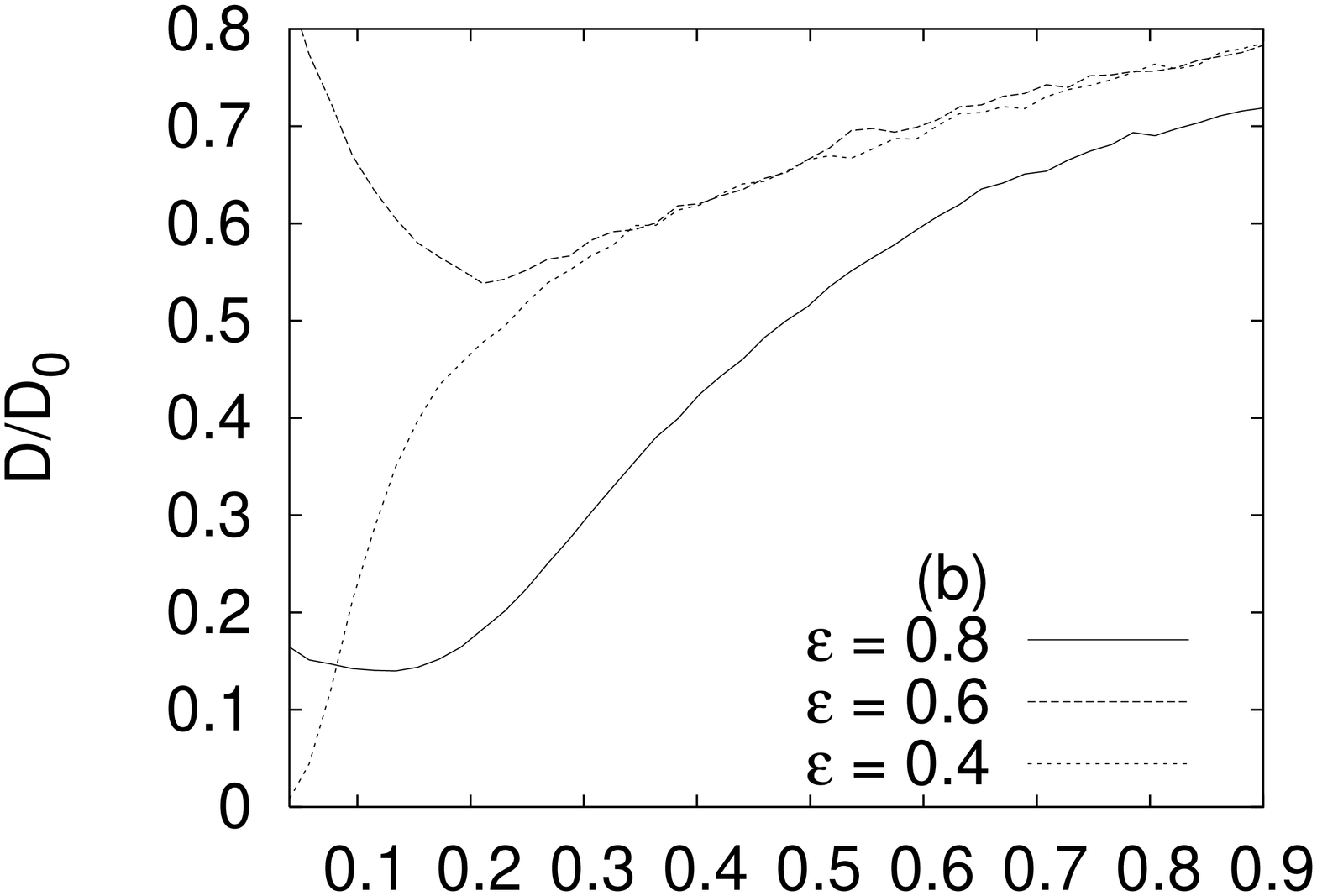}
\includegraphics[height=3.5in,width=2.0in,angle=270]{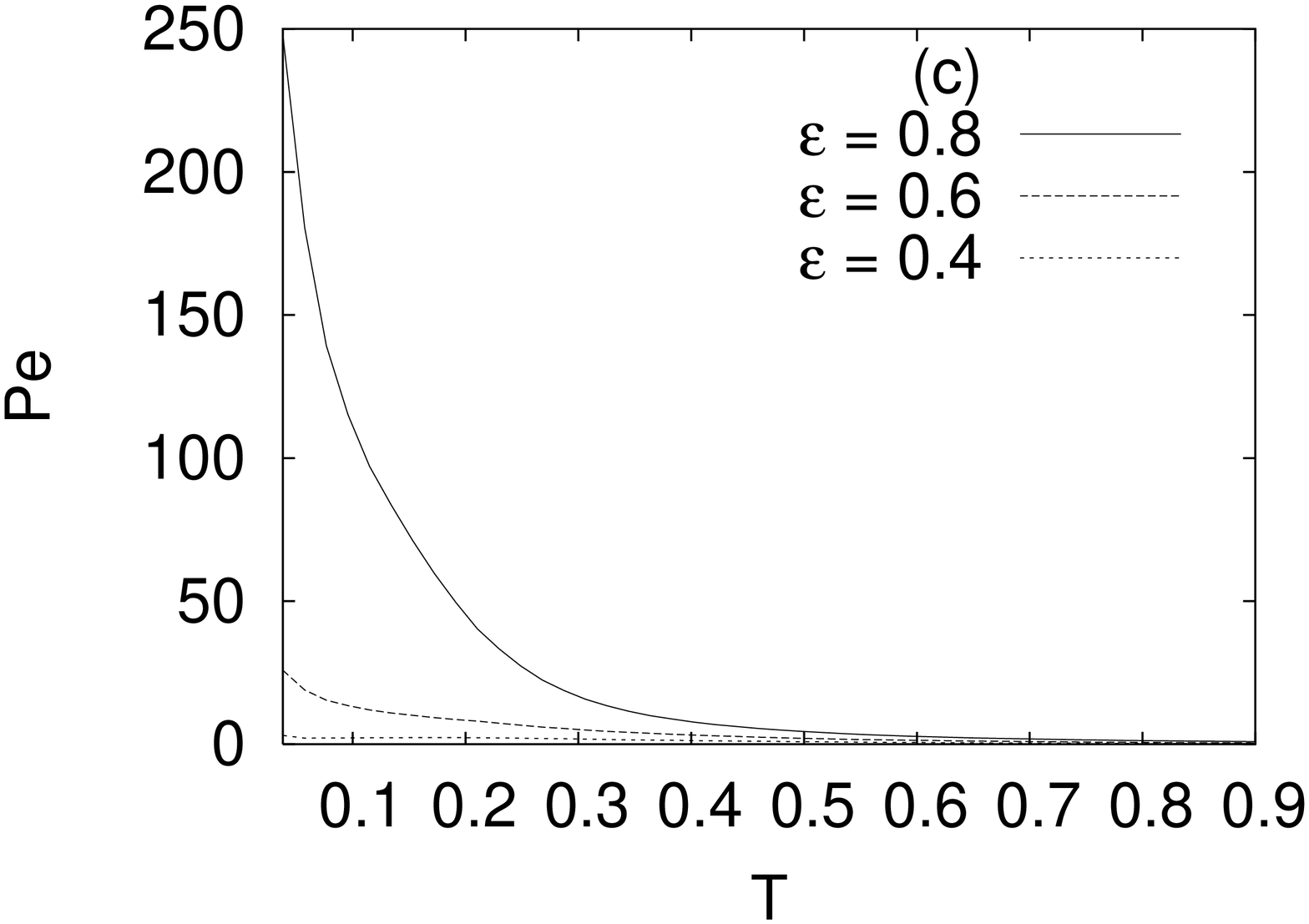}
\caption{Plot of (a) $j$ (b) $D$  and 
(c) $Pe$ (from above) versus $T$ at $F_0 = 0.3$ for 
various values of $\epsilon$ in the temporally asymmetric driving case, 
in the adiabatic limit ($\tau=1000$).}
\label{st}
\end{center}
\end{figure}}

\newcommand{\STI}{
\begin{figure}[htbp]
\begin{center}
\includegraphics[height=3.5in,width=2.0in,angle=270]{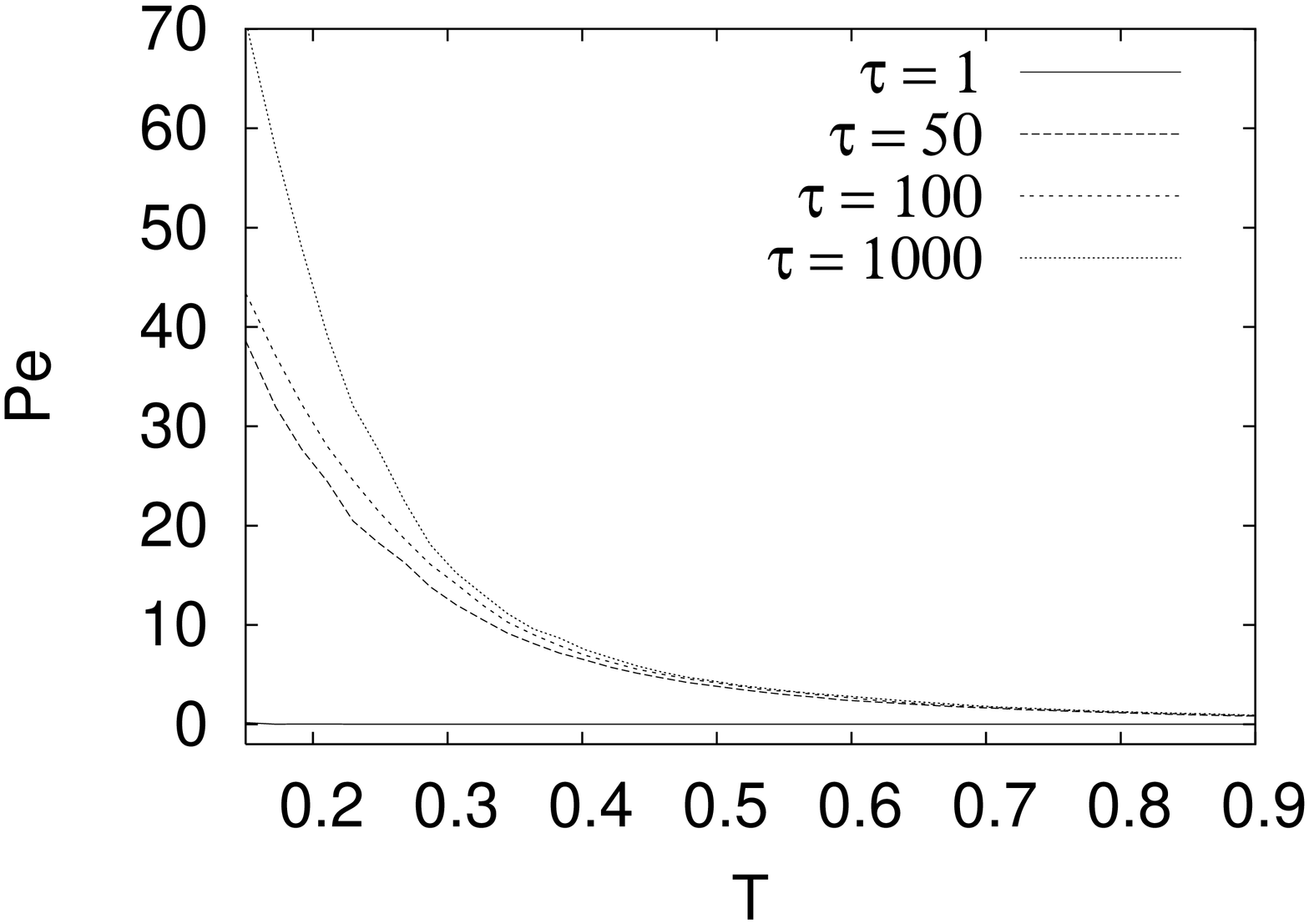}
\caption{Plot of $Pe$ versus $T$ at $\epsilon = 0.8$ and 
$F_0 = 0.3$ for various values of the time period ($\tau$) in the 
temporally asymmetric driving case.}
\label{sti}
\end{center}
\end{figure}}

\begin{document}
\title{Giant coherence in driven systems}
\author{Soumen Roy$^*$, Debasis Dan$^+$ and A. M. Jayannavar$^\dagger$}
\address{${}^{*,\dagger}$ Institute of Physics, Bhubaneswar 751005, India\\${}^+$ Dept. of Biology and Center for Genomics and Bioinformatics, Indiana University, Bloomington 47405 USA}
\eads{\mailto{$^*$ sroy@iopb.res.in}, \mailto{${ }^+$ ddan@indiana.edu}, \mailto{${ }^\dagger$jayan@iopb.res.in}}

\begin{abstract}

We study the noise-induced currents and reliability or 
coherence of transport in  two different classes of rocking ratchets. For 
this, we consider the motion of
 Brownian particles in the over damped limit in both adiabatic and non-adiabatic
 regimes subjected to unbiased temporally symmetric and asymmetric periodic 
driving force. In the case of a time asymmetric driving, we find that even 
in the presence of a spatially symmetric simple sinusoidal potential, highly
 coherent transport occurs.  These ratchet systems exhibit giant 
coherence of 
transport in the regime of parameter space where unidirectional currents in 
the deterministic case are observed. Outside this parameter range, i.e., when 
current  vanishes in the deterministic regime, coherence in transport is very 
low. The transport coherence decreases as a function of temperature and is a non-monotonic 
function of the amplitude of driving. The transport becomes unreliable as we go from the 
adiabatic to the non-adiabatic domain of operation. 

\end{abstract}
~
\maketitle

\section{Introduction}
\label{sec:intro}
Ratchets, Brownian motors or rectifiers are nonequilibrium systems that 
rectify fluctuations in the medium to achieve directed motion
~\cite{reimann,ajdari,apa,amj,rkamj}. The main criteria for these systems 
are spatially extended periodic structures and unbiased external fluctuations 
that drive  the system out of equilibrium. Preferential directed motion is 
 possible if either the potential and/or the external fluctuations is 
asymmetric (broken symmetry)~\cite{reimann}. 
Even in the presence of a spatially asymmetric potential, the principle of 
detailed balance prohibits any net unidirectional current  at equilibrium. 
Only when the system is driven out of equilibrium, this principle 
no longer holds and the Brownian particles can achieve directed motion by 
rectification of thermal fluctuations. These {\it ratchet models} are found 
to have wide ranging applications in physical and biological systems
~\cite{reimann,ajdari,apa,amj,rkamj}.

Considerable amount of work has been devoted to understand the nature of 
 currents and their reversals in different classes of ratchet models (namely 
flashing ratchets~\cite{ajdari}, rocking ratchets~\cite{magnasco}, frictional
 ratchets~\cite{pareek}, etc). Moreover, these ratchets or motors are
 engines at the molecular scale converting input energy from a  
nonequilibrium environment 
into useful work. Hence a lot of attention has been given to the performance 
characteristics of these systems, namely thermodynamic~\cite{sekimoto,parrondo,jstat05,makhnowski} 
and generalised~\cite{generalised1,generalised2} efficiencies. In recent years, 
another important property of these systems is being explored, namely the 
reliability or coherence of transport.

The unidirectional current  of Brownian particles in stochastic ratchets, 
however, is always accompanied by a diffusive spread (dispersion). 
This spread is 
intimately related to the question of reliability or quality of transport  
to the extent that it may completely overshadow the ratcheting effect in a 
system with finite spatial extensions. For example, if a particle on an
 average moves a distance $L$ due to it's finite average velocity, $v$, 
there will always be an accompanying diffusive spread. If this spread is 
much smaller than the distance traveled, then the motion of the particle 
is considered as coherent or reliable. This, in turn, can be quantified 
in terms of a dimensionless number called the P\'eclet number($Pe$), 
which is the ratio of the average velocity, $v$, to the diffusion 
constant, $D$. More specifically $Pe = vL/D$.  In our studies, we take 
$L$ to be the length of the period of the relevant spatially periodic 
potential. Quantitatively, if $Pe\gg2$,  the transport is said to be 
coherent, otherwise it is incoherent or 
unreliable. There exist very few studies, which address the question
 of reliability of transport. $Pe$ for some models of
 flashing and rocking ratchets were found to be $ \sim 0.2$ and $ \sim 0.6$ 
respectively~\cite{low}, implying a less reliable transport. A study on 
symmetric periodic potentials along with a spatially modulated white noise 
showed a coherent transport with $Pe$ less than $3$. In the same 
study a special kind of strongly asymmetric potential was found to increase 
$Pe$ to $20$  in some range of physical parameters~\cite{sch,machura}.
Experimental studies in  biological motors show them to exhibit highly 
efficient and 
reliable transport with $Pe$ ranging from 2 to 6~\cite{block}.  
  In a very recent work, the collective effects of coupled Brownian motors 
were found to show high transport coherence~\cite{bao}. Reliability of 
transport has also been studied in frictional ratchets and coherent 
transport is observed in a part of the parameter space~\cite{frictional}.

In the present work, we study the transport coherence in two different classes 
of rocking ratchets. In the first case (hereafter referred to as {\it case 1}), we study  a ratchet model, where the potential is simply sinusoidal(spatially 
symmetric) while the driving is temporally asymmetric. In Refs.~\cite
{chialvo,mm-amj,rock-prl} an unbiased discontinuous temporally asymmetric  
driving has been considered.  For the case of the asymmetric drive, 
characterised by an additive Poissonian white shot noise with a constant 
bias ensuring zero time average, analytical solutions have been obtained 
for the noise induced currents~\cite{poisson}.
The time asymmetric drive can be generated by the 
application of biharmonic drive at frequencies $\omega$ and $2\omega$. This 
phenomenon is known as harmonic mixing~\cite{marchesoni} and has been studied 
extensively in the context of ratchet dynamics~\cite{savelev}, in the problem 
of kink-assisted directed energy transport in soliton systems~\cite{chacon} etc.  Experimentally time asymmetric ratchet mechanism has been used to generate 
photo-current  in semiconductors~\cite{shmelev} (for details see section 5.2 
of Ref.~\cite{reimann}). Recently, Brownian motors with time-asymmetric 
driving in 
a periodic potential have been realised in cold atoms in a dissipative optical 
lattice~\cite{schiavoni}. 
In the second case (hereafter referred to as {\it case 2}), the 
ratchet is characterised by a spatially asymmetric potential driven by a 
temporally symmetric ac force. We report our results on the reliability of 
transport on the model earlier studied by Bartussek et al~\cite{bartussek} 
in the same parameter space explored by them.
  Our work on transport coherence is relevant to the aforementioned 
experimental studies~\cite{apa, shmelev, schiavoni}. One can readily perform 
measurements of transport coherence in experimental set-ups akin to Ref.
~\cite{schiavoni}.
 We show throughout 
this work that that these ratchets exhibit a {\it generic effect}  
in the deterministic limit(absence of noise or temperature) : if the ratchet 
exhibits a finite current ,  one observes giant coherence at low 
temperatures while if the current  vanishes, the associated transport 
coherence is very low. Moreover, this enhanced coherence is maintained as 
long as currents in the backward direction are suppressed. The suppression 
of backward currents also leads to an enhanced thermodynamic 
efficiency of energy transduction ~\cite{jstat05,makhnowski} in 
absence of which the thermodynamic
 efficiency in ratchet systems is very low~\cite{kamegawa}.
The transport coherence decreases as a function of 
temperature and is a non-monotonic function of the driving amplitude. 
Moreover, the transport becomes less reliable as we approach the 
non-adiabatic domain of operation.
\FORCE
\section{Model:}
\subsection{Case 1 : spatially symmetric potential with temporally 
asymmetric driving}
The starting point of our equation is the Brownian motion of an overdamped  
particle in presence of a potential and a driving force which can be 
described by the overdamped Langevin equation \cite{risken}.
\beq
\gamma\dot{x} = -\partial_x[V(x)-xF(t)] + \xi(t)
\label{laneqn}
\enq
 The thermal noise is modeled by a zero mean Gaussian white noise $\xi(t)$,
 with correlation $\langle \xi(t)\xi(t')\rangle = 2k_BT\gamma\delta (t-t')$.
The periodic potential is chosen as $V(x) = V_0~sin(x)$. Since $V(x)$ is 
symmetric, to generate unidirectional currents, one has to apply a 
time asymmetric driving. $F(t)$ is the externally applied 
time periodic driving force, whose average over a time period is zero
~\cite{jstat05,chialvo,mm-amj}
 and is given by 
\begin{eqnarray}
F(t)&=& \frac{1+\epsilon}{1-\epsilon}\, F_0,\,\, (n\tau \leq t < n\tau+ \frac{1}{2}
 \tau (1-\epsilon)), \\ \nonumber &=& -F_0,\,\, (n\tau+\frac{1}{2} \tau(1-\epsilon)
 < t \leq (n+1)\tau),
\end{eqnarray}
Here, the parameter $\epsilon$ signifies the temporal asymmetry in the 
periodic forcing while $\tau$ is the time-period and $n = 0, 1, 2, \dots$ 
is an integer. The force profile is shown in Fig. \ref{forceee}. 

\subsection{Case 2 : spatially asymmetric potential with temporally symmetric 
driving}

We consider the same ratchet model as considered by Bartussek 
et al~\cite{bartussek} with the potential 
 $V(x)=-\frac{V_0}{k} [sin(kx)+0.25~sin(2kx)]$ with $k=2\pi$. 
 We now impose a periodic unbiased ac 
force, $F(t) = F_0 sin(\omega t)$. The underlying asymmetric potential 
breaks the symmetry of the system and generates a current .

\section{Numerical details}

The analytical expressions for the currents($j$) and diffusion 
coefficient($D$)  can only
 be obtained in the adiabatic or quasi static limit , i.e., when the frequency 
of the driving force is small compared to the other frequency scales in the 
problem~\cite{jstat05,harada}. In such a situation, the system can be 
considered to be in a steady state at each instant of time. For the 
general case, we are forced to take recourse to  numerical 
simulations~\cite{toral,nm}. In this work, we have used Langevin 
simulations to evaluate $j$ and $D$.
 We use the Huen's method in these simulations\cite{toral} and calculate the 
current in the asymptotic regime. The expressions for $j$ and $D$ are given by,
\beq
j = \left \langle \frac{x(t)-x(t_0)}{t-t_0} \right \rangle
\label{$j$ eq}
\enq

and 

\beq
D = \lim_{t \to \infty}
\frac{1}{2t} \left [ \langle {x^2(t)}  \rangle - {\langle x(t) \rangle} ^2 \right]
\label{diffeq}
\enq

Here $\langle \dots \rangle$ denotes ensemble averaging. We discard the 
initial transients ($t_0 = 500 \tau$) and then evolve the system for 
$t = 25000 \tau$. In each case the time-step is taken equal to $0.01$ and    
  the averaging is done over $5000$ ensembles.
 
In all the figures to follow,
the physical quantities taken are in dimensionless units~\cite{machura}. 
Energies are scaled with respect to the potential strength, $V_0$; 
lengths are scaled with respect to the spatial period of the  potential.  Also, 
the frequency of oscillation is scaled with respect to the friction coefficient 
 and $F_0 \equiv \frac{FL}{V_0}$. As a check we have reproduced the main results 
of Refs.~\cite{bartussek,haken,hanggi-prl-2001}

\section{Results and Discussion}
\subsection{case 1}
\label{results}
\SF

Fig.~\ref{sf} shows the variation of $j$, $D$  and
 $Pe$ versus force. In our work, current($j$)  and velocity($v$) carry 
the same meaning. We have taken the time period $\tau = 1000$ 
to be very large so that we are in the quasi static limit. For a simple
  potential $V(x) = V_0~sin~x$ in the presence of a static force
($F_0$), in the deterministic limit current  flows only when $F_0$ crosses
 a critical
 threshold($F_c$), namely, $F_0 > F_c = 1$. Beyond the critical threshold, 
barriers to the motion in the forward direction disappear. Consequently, the
particle is in the running state (i.e., the particle is free to move). Below
 the critical field, the particle experiences barriers in the direction of
 the applied field and hence at temperatures $T \to 0$, particle will be
trapped at a local minimum of the potential (i.e., in a locked state). It is 
also known that {\it giant diffusion} arises around the ``dynamical 
bottleneck" at $F_0 = F_c = 1$  and is expected as a fallout of the 
instability between the 
locked state and the running state~\cite{hanggi-prl-2001,danpre66,schreier}. 
The peaks in the $D$  versus force curve around 
$F_c \approx 1$, gets sharpened as the temperature is reduced. 
At high temperatures, due to thermal smearing, the peaks become 
broader. 
 In the adiabatic limit, we can consider the total current  to arise from 
the sum of the contributions of the fraction of the current  when the field 
is in the forward direction and the fraction of the current  when the field 
is in the backward direction~\cite{frictional,harada}. 
In the same limit, we can also consider the total diffusion coefficient  to 
arise from the sum of similar contributions of the diffusion coefficients  
from force fields in the forward and backward directions. As we increase the 
amplitude of the temporal force $F_0$, in the deterministic limit 
($T \to 0$ limit), current  in the forward direction starts flowing when 
$\frac{1+\epsilon}{1-\epsilon} F_0 > 1$ or $F_0 > \frac{1-\epsilon}{1+\epsilon}$.  
We have chosen $\epsilon = 0.8$, hence, one observes significant currents only 
above $F_0 > 0.11$. As we increase the amplitude $F_0$,
 $j$  increases till $F_0$ becomes of the order of $1$. 
Up to this limit, current  in the backward direction is absent (as the force
 applied in the backward direction is $F_0$ which is independent of $\epsilon$).
Thus in the range of $F_0$ between $\frac{1+\epsilon}{1-\epsilon}$ and $1$, the
 current  increases monotonically. Beyond $F_0 > 1$ the barriers to motion for 
a particle in both directions disappear and consequently 
 current  decreases as we increase $F_0$ further. 
The temperature only broadens the peak and the value of $F_0$ at
which the peak appears, shifts to the left. This is because, temperature can 
facilitate current  in the backward direction, even when barriers are present.

In fig.~\ref{sf}(b), we have plotted $D$  (scaled with 
respect to the bare diffusion coefficient, $D_0\equiv k_BT/\gamma$) versus 
the driving force $F_0$. For very small values of the driving force, i.e, 
when $F_0 \ll 1$, $D \ll D_0$  
 due to the presence of barriers in motion in both directions. Two peaks 
are observed at $F_0 \approx 0.1$ and $1$, 
which correspond to the vanishing of barriers for forward and backward 
directions respectively (i.e., instability points) as discussed earlier. The 
diffusion peak around $F_0 = 1$ is pronounced and has a value greater than $1$,
 i.e., $D \gg D_0$. 
This is an anticipated effect~\cite{hanggi-prl-2001,danpre66}. 
The peak broadens with the rise in temperature. However, unlike the peak 
in $j$, it does not shift with temperature~\cite{hanggi-prl-2001}.

We notice clearly from fig.~\ref{sf}(a) and fig.~\ref{sf}(b) that in the range 
 between $F_0 \approx 0.1$ and $1$,  {\it enhanced currents are accompanied 
by minimal diffusion}. As a consequence it is in this region, that one observes 
enhanced or giant transport coherence ($Pe \approx 450$ for $T = 0.05$ around 
$F_0 \approx 0.6$). The observed values are very much larger than those 
obtained for other ratchet systems~\cite{low,sch,machura,frictional}.  
 It may be noted that, in the regime of giant coherence, current  in the 
backward direction is suppressed as mentioned earlier. Precisely in this 
regime of $F_0$, it has been shown that the thermodynamic  efficiency
($\eta$)~\cite{jstat05} and the generalised efficiency~\cite{generalised3} 
is quite high, even though the ratchet operates in an irreversible mode.

\NONAD

We now very briefly discuss the nature of $j$, $D$ and $Pe$ 
 as a function of $F_0$ in the non-adiabatic limit. 
For this, we have plotted in fig.~\ref{nonad} the variation of  
$j$, $D$  and $Pe$ versus $F_0$ for 
$\tau = 5$ at $T = 0.2$ and $\epsilon = 0.8$. We notice that $j$  exhibits
 a peak shifted to the right as compared to the graph in the adiabatic limit
 i.e., fig.~\ref{sf}(a).  The currents are very low for small $F_0$,  
even for the regime around ($F_0 \geq 0.11$). In this 
regime, particle cannot take advantage of vanishing of barriers in the forward 
direction as it will not be able to traverse a distance of half a period in 
the duration in which the force is in positive direction, i.e., force reverses 
its sign before the particle could traverse a distance of half a period. 
 However, on increasing the value of $F_0$, the particle will 
naturally take advantage of the vanishing barriers. Hence, peak shifts 
towards the right. Unlike adiabatic case, $D$  does not 
exhibit a two-peak structure. The peak at smaller value of $F_0 \approx 0.11$ 
disappears. Here too the particle does 
not take advantage of the vanishing of barriers in the forward motion. $Pe$ 
 exhibits values which are 
very much smaller than those obtained in the adiabatic limit. Hence, coherence 
in transport is reduced as the time-period is reduced.

\ST

Fig.~\ref{st} shows the variation of $j$, $D$, and 
$Pe$ with respect to $T$ (scaled with respect to $V_0$, the strength
 of the potential) for various values of $\epsilon$ at $F_0 = 0.3$ .  
The $j$  and $D$   versus $T$ curves show the crucial 
role played by the temporal 
asymmetry factor $\epsilon$. The higher values of current  are obtained for 
higher $\epsilon$. For $\epsilon = 0.4$ at $F_0 = 0.3$, barriers for
the motion of the particle are present in both the directions and as a result 
the current  vanishes in the zero temperature limit. Thus, for intermediate 
values of temperature, a peak is witnessed. For other values of $\epsilon$, 
namely, $\epsilon = 0.6$ and  $\epsilon = 0.8$, barriers to the motion in 
the forward direction vanish but are present in the backward direction. 
Hence, at zero temperature, we get a finite current  which 
vanishes at high temperature. For $\epsilon  = 0.8$ the current  decreases 
monotonically whereas for $\epsilon = 0.6$ the current  exhibits a small peak.

The origin of the temperature axis in fig.~\ref{st}(b) is at $T = 0.04$. 
The scaled diffusion coefficient  ($D/D_0$) exhibits a minima as a function 
of temperature in the range considered (for $\epsilon = 0.6, 0.8$). This is 
due to the fact that the scale $D_0 \equiv k_BT/\gamma$. We have observed that 
without this scaling factor, $D$ increases 
monotonically with temperature, starting at zero at $T = 0$. In the high 
temperature limit, $T > 1$, (i.e., when $T \gg V_0$),  $D/D_0 \to 1$ as 
anticipated.  At low temperatures, $D/D_0$ exhibits a non-monotonic 
behaviour as a function of $\epsilon$. 
Other quantities like the thermodynamic efficiency~\cite{jstat05} 
and the generalised efficiency~\cite{generalised3} also exhibit a 
non-monotonic behaviour as a function of the temporal asymmetry 
factor $\epsilon$. This is not 
surprising as there are two competing effects : as one increases $\epsilon$, 
the barriers in the forward direction are reduced while the fraction of the 
time period during which the particle is subject to a positive force is also 
decreased.

From fig.~\ref{st}(c) we see that $Pe$ diminishes as we increase the 
temperature. Higher the $\epsilon$ value, higher is the coherence.  This 
enhanced coherence is sustained over a large temperature regime. We would 
like to emphasise that at very low temperatures ( $T \to 0$), finite current  
results for a range of parameters. However, $D$ tends to zero in the same 
range. As a result, $Pe$ exhibits a divergent behaviour. Hence, to avoid
 numerical errors, the origin of the temperature axis is chosen as $T = 0.04$. 
It should be noted that for $\epsilon=0.4$, 
current  in the deterministic limit vanishes as can be observed in 
fig.~\ref{st}(a) and consequently transport 
coherence is very low as seen from fig.~\ref{st}(c). 
These results bring out the generic effect mentioned 
in Section~\ref{sec:intro}.

\STI

In fig.~\ref{sti}, we show the variation of $Pe$ for a fixed value of 
$\epsilon =  0.8$, as a function of time-period. The origin of the 
temperature axis is at $T = 0.15$. It is clear, that the transport which is 
coherent in the adiabatic limit ($\tau = 1000$) loses its coherence 
as the  non-adiabatic limit is reached. This conclusion about the superior 
reliability of  transport of the ratchet at the adiabatic limit is 
generally true(we have verified this separately).

\subsection{case 2}

Now we turn to case 2 , which has been studied extensively for the nature 
of currents by Bartussek et al \cite{bartussek}.
\HBF
In fig.~\ref{hbf}, we have plotted $D$ and $Pe$ as a function of $F_0$ at 
$T = 0.1$ for various frequencies $\omega$. In 
fig.~\ref{hbf}(a) we reproduce the same results as obtained in fig. 1(b) of Ref.
~\cite{bartussek}. $\omega = 0.25$ corresponds to the adiabatic regime. Here, 
current  flows in a positive direction and exhibits a peak. During the fraction 
of the time-period when force is in positive direction, the particle 
experiences 
a smaller barrier in the forward direction as opposed to the fraction of the 
period when force is in the negative direction and particle experiences a 
higher potential barrier. During each half cycle particle traverses a 
distance much larger than the spatial period of the potential. In this
 regime, the particle takes advantage of the presence of anisotropy in the 
potential and hence positive current  arises~\cite{reimann}. As we approach 
the non-adiabatic limit ($\omega = 7$, $\omega =  10$), we observe multiple 
current  reversals. For details see Refs.~\cite{bartussek,danpre63}. 

In fig~\ref{hbf}(b), we have 
plotted $D$  as a function of $F_0$ for various 
frequencies. $D/D_0$ starts  from a finite value at $F = F_0$, and 
asymptotically approaches $1$  as expected. Some local 
maxima are observed at finite driving frequencies. These features are also 
seen for the case of symmetric periodic potential driven by a temporally 
symmetric periodic force as discussed in Refs.~\cite{haken,schreier}. 
These peaks are attributed to optimised enhancement of the escape rate by 
modulation for a given noise strength. As discussed in Refs.~
\cite{haken,schreier}, for certain values of the driving force (or forces), 
the position probability peak of the particles may just happen to be on the 
top of the potential barrier and the diffusion is naturally more than there 
would have been if this peak had been located elsewhere (especially if it  
was at the potential minimum for example). 

In fig.~\ref{hbf}(c) we have plotted corresponding $Pe$ as a 
function of $F_0$.  We readily notice that even in the 
adiabatic limit, transport is incoherent since values of $Pe \approx 0.45$ 
are obtained which further decrease as we cross over to the 
non-adiabatic limit . Therefore the noise induced transport in this system is 
completely incoherent in the range of parameters considered here.  {\it 
This range corresponds to the current  being zero in the deterministic regime.}

We would like to emphasise that the phenomenon of current  reversal in  
ratchets plays a major role in devising novel separation techniques for 
nanoparticles~\cite{amj}. Once the current  reversal as a function of 
any parameter is established, it follows 
readily that current  reversals can be observed by varying other parameters in 
the system~\cite{reimann,rkamj}. In these devices, particles with different 
masses move in the opposite direction which can be readily separated. However, 
we notice from the figure that, around the current  reversals, the P\'eclet 
numbers are quite small and transport is incoherent.

 In fig.~\ref{hbfspl} we plot the variation of $Pe$ as 
a function of $F_0$ in the adiabatic limit($\omega~=~0.25$) for two fixed 
values of $T=0.01$ and $T=0.1$. The inset 
shows the variation of $j$  as a function of $F_0$ at $T=0.01$ and 
$T=0.1$. At $T=0.01$, we are close to the deterministic regime, 
where the values of $j$  and 
$Pe$ are expected to be large.  The figure corresponding 
to $T=0.01$ shows a higher value of $Pe$. 
In the adiabatic limit the total current  
is expected to be the sum of the current  contributions due to the forward 
and backward driving. It is readily  seen from the figure, that  
the value of $Pe$ in the curve for which $T=0.01$, is 
high above $F_0~\approx~0.72$, i.e, when the barriers to the 
current(inset) in the forward direction are absent. The current  
steadily increases with driving until $F_0~\approx~1.5$ where a peak in 
current  is observed.  However, beyond this value of $F_0$,  the barriers to 
the motion in the other direction also disappear and hence the net current 
starts decreasing as can be seen from the inset of fig~\ref{hbfspl}. 
It is notable that the value of $Pe$ in the observed regime 
beyond the value of  $F_0~\approx~0.72$ is $\gg~2$ and in fact seen to 
be as high as $16$. In the region $F_0~<~0.72$, where the current  in 
the low $T$ limit vanishes, the transport is incoherent. 
For $T=0.1$, the particles can take the aid of significant thermal 
fluctuations to cross the barriers in both directions. Hence, the current  
values at $T=0.1$ for $F_0~>~0.72 $ are much lower than those observed for 
$T=0.01$. Therefore, the associated $Pe$ 
values are much lower for $T=0.1$ and the transport is seen to be 
completely unreliable in fig.~\ref{hbfspl}.

\HBFSPL

\HBTSPL

 In fig.~\ref{hbtspl} we plot the variation of $Pe$ 
and $j$ (inset) as a function of $T$ at two fixed values of 
$F_0=0.9$ and $F_0=1.25$. We have restricted ourselves to the 
adiabatic($\omega~=~0.25$) domain of operation, where the values of 
$j$  and $Pe$ are expected to be large. For these values of $F_0$, a finite 
current  results at $T=0$, as shown in the inset of fig.~\ref{hbtspl}. Hence, 
giant coherence is expected at low temperatures. To avoid divergence of 
$Pe$, the origin of the $T$ axis is chosen at $T=0.0025$. With increase in $T$, the transport coherence diminishes.
\HBT

In fig.~\ref{hbt}, we have plotted $j$, $D$ and $Pe$ as a function of $T$ for 
various frequencies mentioned in fig~\ref{hbt} at $F_0 = 0.5$. 
Fig.~\ref{hbt}(a) reproduces 
the currents of fig. 1(b) of Ref.~\cite{bartussek}. In the adiabatic limit 
 ($\omega = 0.25$), current  remains positive and exhibits a peak. Current in 
the non-adiabatic limit ($\omega = 4.0$ and $\omega = 7.0$) starts with a 
negative value and exhibits current  reversals ~\cite{bartussek,danpre63}. 
$D$  saturates to
 a value $D/D_0 \approx 1$ in the high temperature limit. Depending on the 
temperature regime, whether $D$ is a monotonic or non-monotonic function of 
frequency can be inferred from fig.~\ref{hbt}(b). Corresponding $Pe$  
are plotted in fig~\ref{hbt}(c). Similar to the behaviour seen in 
fig.~\ref{hbf}(c), we observe that as we go from 
the adiabatic to the non-adiabatic regime, transport becomes incoherent.  
Beyond $T = 0.1$, the transport becomes  completely unreliable ($Pe \ll 2$). 
\section{Conclusions}
We have studied the Brownian dynamics of a particle in a symmetric sinusoidal 
potential in presence of time-symmetric unbiased forcing. 
We have shown that the resulting
 fluctuation induced currents exhibit giant coherence in transport. 
This is observed 
in the parameter space where currents are finite in the deterministic limit. 
Moreover, this 
coherence can be sustained over a large temperature range and variations in 
other relevant physical parameters. Transport is most coherent in the 
adiabatic limit and decreases as we approach the non-the 
adiabatic limit. 
 The coherence in transport reduces as a function of temperature and is a 
 non-monotonic function of the amplitude of driving.

In general, the ratchet systems which favour currents in the forward 
direction and suppresses currents in the backward direction are expected to 
show enhanced coherence, other examples being flashing ratchets where two 
periodic states are displaced with respect to each other~\cite{makhnowski}. 
The ratchet systems at finite driving frequencies studied in this work exhibit 
several complex features in the nature of current  and diffusion coefficient 
in the deterministic limit. Current exhibits quantisation (plateaus) associated 
with phase or frequency locking behaviour as a function of the amplitude of the
 driving force and other parameters~\cite{bartussek,danpre63,danmenon,reguera}. 
Correspondingly, diffusion exhibits giant peaks and crests in presence of small 
noise. These curves develop oscillatory  
features (namely resonances and antiresonances) in the presence of small noise 
 (multiple peaks in diffusion and currents can be observed). These intriguing 
features can be attributed to the complex dynamics of the particle which arises 
due to the combined effects between non-linearity, frequency of driving and
 noise~\cite{reguera}. 
However, all of these complex features are not robust in the presence of 
finite noise, the subject matter of which is under study vis-a-vis the 
coherence in transport.\\

\noindent
{\bf Acknowledgement}\\

One of the authors(AMJ) thanks M.C. Mahato for useful discussion
 and helpful suggestions.\\

\noindent
{\bf References}\\

 \end{document}